\documentclass[                     
               showpacs,            
               preprintnumbers,     
               aps,                 
               prd,                 
               superscriptaddress,      
               nofootinbib,         
               tightenlines,        
               floats,floatfix      
               ]{revtex4}

\usepackage{amsmath}
\usepackage{amssymb}
\usepackage{graphicx}
\usepackage{color}

\newcommand{\Real}{\mathbb{R}}

\newtheorem{lemma}{Lemma}
\newcommand{\proof}{\noindent {\bf Proof. }}
\newcommand{\qed}{\hfill $\fbox{\hspace{0.3mm}}$ \vspace{.3cm}} 

\begin{document}


\title{On the instability of charged wormholes supported by a ghost scalar field}


\author{J. A. Gonz\'alez}
\affiliation{Instituto de F\'{\i}sica y Matem\'{a}ticas, Universidad
              Michoacana de San Nicol\'as de Hidalgo. Edificio C-3, Cd.
              Universitaria, 58040 Morelia, Michoac\'{a}n,
              M\'{e}xico.}

\author{F. S. Guzm\'an}
\affiliation{Instituto de F\'{\i}sica y Matem\'{a}ticas, Universidad
              Michoacana de San Nicol\'as de Hidalgo. Edificio C-3, Cd.
              Universitaria, 58040 Morelia, Michoac\'{a}n,
              M\'{e}xico.}

\author{O. Sarbach}
\affiliation{Instituto de F\'{\i}sica y Matem\'{a}ticas, Universidad
              Michoacana de San Nicol\'as de Hidalgo. Edificio C-3, Cd.
              Universitaria, 58040 Morelia, Michoac\'{a}n,
              M\'{e}xico.}


\date{\today}


\begin{abstract}
In previous work, we analyzed the linear and nonlinear stability of static, spherically symmetric wormhole solutions to Einstein's field equations coupled to a massless ghost scalar field. Our analysis revealed that all these solutions are unstable with respect to linear and nonlinear spherically symmetric perturbations and showed that the perturbation causes the wormholes to either decay to a Schwarzschild black hole or undergo a rapid expansion.
Here, we consider charged generalization of the previous models by adding to the gravitational and ghost scalar field an electromagnetic one. We first derive the most general static, spherically symmetric wormholes in this theory and show that they give rise to a four-parameter family of solutions. This family can be naturally divided into subcritical, critical and supercritical solutions depending on the sign of the sum of the asymptotic masses. Then, we analyze the linear stability of these solutions. We prove that all subcritical and all critical solutions possess one exponentially in time growing mode. It follows that all subcritical and critical wormholes are linearly unstable. In the supercritical case we provide numerical evidence for the existence of a similar unstable mode.
\end{abstract}




\maketitle


\section{Introduction}

Wormhole spacetimes in Einstein's gravitational theory have received considerable attention in the literature. Presumably, this is due to their interesting topological and causal properties which open the door to spectacular phenomena such as interstellar travel and time machines \cite{mMkT88,mMkTyU88,vFiN90}. More recently, wormholes have also been proposed as black hole mimickers \cite{tDsS07} (see also \cite{nMtZ08}). However, there are several problems which make these solutions somehow unattractive from a physical point of view and pushes them to the verge of science fiction. First of all, wormholes need to be supported by exotic matter if they are to be considered as asymptotically flat, globally hyperbolic spacetime solutions of Einstein's field equations \cite{mMkTyU88,jFkSdW93}. This means that they require an energy-stress tensor which violates the (averaged) null energy condition, a phenomena which has not been observed for classical matter fields in the laboratory. Furthermore, to our knowledge, there are no static, asymptotically flat wormhole models which have been shown to be linearly stable with respect to {\em arbitrary} linear fluctuations of the metric and matter fields in the given model. This is to be contrasted with static, asymptotically flat black holes with a regular horizon in vacuum or electrovacuum spacetimes, which are known to be stable with respect to linear perturbations \cite{tRjW57,fZ70,bKrW87,vM74a,vM74b,vM75}. Therefore, even if some exotic form of matter could be found in the Universe, it is not clear whether or not it could be used to form static wormholes.

In previous work \cite{jGfGoS09a,jGfGoS09b} we analyzed the question of wormhole stability for a very simple matter model which consists of a massless ghost scalar field, that is, a massless scalar field whose kinetic energy has a reversed sign. We found that {\em all} static and spherically symmetric wormholes in this theory are unstable with respect to linear and nonlinear perturbations. Each of these wormholes possesses a single unstable mode which causes the wormhole to collapse to a black hole or to undergo a rapid expansion. Furthermore, the time scale associated to the linear instability is of the order of the areal radius of the wormhole's throat divided by the speed of light.

The purpose of this article is to analyze whether or not one could stabilize these wormholes. One possible mechanism for stabilization is to consider stationary, rotating generalization of such wormhole solutions and hope that they become stable if the angular momentum is large enough. Slowly rotating wormholes have been constructed in \cite{pKsS08} by studying linear perturbations of the static solutions. However, such slowly rotating solutions cannot be expected to be stable since the unperturbed solutions are unstable. Wormhole solutions of the full nonlinear field equations which represent rotating generalizations of the static ones have been considered in \cite{tMdN06,tM09}, but a detailed stability analysis of such solutions is not expected to be simple. For this reason we consider, here, a different possible stabilization mechanism which is the addition of an electromagnetic charge. Therefore, we add a Maxwell field to the ghost field and consider static, spherically symmetric wormhole solutions in this theory. It turns out that the resulting field equations can be integrated exactly and give rise to a four-parameter family of wormhole solutions. This is described in section~\ref{Sec:StaticSolutions}. Some of these solutions have been found in \cite{tM09}. The phase space of solutions can naturally be divided into subcritical, critical and supercritical wormholes depending on whether the sum of their asymptotic Arnowitt-Deser-Misner (ADM) masses is negative, zero or positive, respectively.

Next, we analyze the stability of such wormholes with respect to linear, spherically symmetric fluctuations. In section~\ref{Sec:Pulsation} we derive the perturbation equations and cast them into a constrained wave system for two gauge-invariant quantities. We then discuss different ways of decoupling this system. In section~\ref{Sec:LinStabAnal} we first prove that all subcritical and critical wormholes are linearly unstable. We do this using the same techniques as in our previous work, namely the theory of Schr\"odinger operators on the real line. Then, we focus our attention to supercritical wormholes which turn out to be more interesting. In this case, we are not able to reduce the perturbation equations to a single wave equation with regular potential. Therefore, we analyze the numerical stability of these wormholes by numerical integration of the constrained wave system. The results indicate that these wormholes are linearly unstable as well, with the perturbations exhibiting an exponential growth of the form $e^{\beta\tau}$ for large values of proper time $\tau$ at the throat, where $\beta > 0$. Interestingly, we find that if the charge is large enough, this exponential growth is accompanied with an oscillating factor of the form $\cos(\omega\tau - \delta)$ for some frequency $\omega$ and phase $\delta$.
Our results also indicate that the growth rate $\beta$ decreases monotonically when the charge increases, and for the range of parameters used in our numerical simulations we observe that $\beta$ can be decreased by a factor of more than $100$ when compared to the uncharged case. Finally, we show by a numerical matching algorithm that these asymptotic solutions correspond to eigenfunctions of the spatial perturbation operator corresponding to a {\em complex} eigenvalue $-(\beta + i\omega)^2$.

A summary of our results and conclusions are given in section~\ref{Sec:Conclusions}. Technical properties of the static wormhole solutions which are needed for the stability analysis are stated and proved in an appendix.

\section{Static, spherically symmetric charged wormholes}
\label{Sec:StaticSolutions}

We consider a spherically symmetric gravitational field which is
coupled to a massless ghost scalar field $\Phi$ and to an
electromagnetic field $F$. Therefore, choosing suitable local
coordinates $t,x,\vartheta,\varphi$, the metric has the form
\begin{equation}
ds^2 = -e^{2d} dt^2 + e^{2a} dx^2 
 + e^{2c}\left( d\vartheta^2 + \sin^2\vartheta\; d\varphi^2 \right),
\label{Eq:SphericalMetric}
\end{equation}
where the functions $d = d(t,x)$, $a = a(t,x)$ and $c = c(t,x)$ depend
only on the time coordinate $t$ and the spatial coordinate $x$. We
assume that $\Phi$ and $F$ are also spherically symmetric which means
that $\Phi = \Phi(t,x)$ and that $F$ has the form
\begin{equation}
F = \alpha\, dt \wedge dx 
  + \beta\,  d\vartheta \wedge \sin\vartheta\; d\varphi,
\label{Eq:SphericalF}
\end{equation}
with two functions $\alpha = \alpha(t,x)$ and $\beta = \beta(t,x)$. We
are interested in traversable wormhole geometries which consist of a
throat connecting two asymptotically flat ends at $x\to +\infty$ and
$x\to -\infty$, respectively. This means that the areal radius $r =
e^c$ is strictly positive and proportional to $|x|$ for large $|x|$
and that the $2$-manifold $(\tilde{M},\tilde{g}) = (\Real^2,-e^{2d}
dt^2 + e^{2a} dx^2)$ is regular and asymptotically flat at $x\to
\pm\infty$.

The equations of motion are
\begin{eqnarray}
&& R_{\mu\nu} = \kappa_0\left[ F_\mu{}^\sigma F_{\nu\sigma} 
 - \frac{1}{4}\, g_{\mu\nu} F^{\sigma\rho} F_{\sigma\rho} \right]
 + \kappa\; \nabla_\mu\Phi \cdot \nabla_\nu\Phi,
\label{Eq:Einstein}\\
&& \nabla^\mu F_{\mu\nu} = 0, \qquad 
   \nabla_{[\sigma} F_{\mu\nu]} = 0,
\label{Eq:Maxwell}\\
&& \nabla^\mu\nabla_\mu\Phi = 0,
\label{Eq:Wave}
\end{eqnarray}
where $R_{\mu\nu}$ and $\nabla_\mu$ denote, respectively, the Ricci
tensor and the covariant derivative associated with the spacetime
metric $g_{\mu\nu}$. In terms of Newton's constant $G$, the coupling constants $\kappa_0$ and $\kappa$ are given by $\kappa_0 = 2G > 0$ and $\kappa = -8\pi G < 0$
which is negative due to the fact that $\Phi$ describes a ghost scalar field. For the spherically symmetric ansatz
(\ref{Eq:SphericalMetric},\ref{Eq:SphericalF}) Maxwell's equations (\ref{Eq:Maxwell}) imply that $\alpha = Q_e e^{a+d-2c}$ and $\beta =
Q_m$ with $Q_e$ and $Q_m$ two constants representing, respectively,
the electric and magnetic charge. Setting $Q:=\sqrt{Q_e^2 + Q_m^2}$
the remaining Eqs. (\ref{Eq:Einstein},\ref{Eq:Wave}) yield the evolution equations
\begin{eqnarray}
\partial_t\left( e^{a-d} a_t \right) - \partial_x\left( e^{d-a} d_x \right)
 - e^{a-d} c_t^2 + e^{d-a} c_x^2 - e^{a+d-2c} 
 &=& -\kappa_0 Q^2 e^{a+d-4c}
     - \frac{\kappa}{2}\left[ e^{a-d}\Phi_t^2 - e^{d-a}\Phi_x^2 \right],
\label{Eq:Ev1}\\
\partial_t\left( e^{a-d+2c} c_t \right) 
 - \partial_x\left( e^{d-a+2c} c_x \right) &=& -e^{a+d}
 + \frac{\kappa_0}{2}\, Q^2 e^{a+d-2c},
\label{Eq:Ev2}\\
\partial_t\left( e^{a-d+2c} \Phi_t \right) 
 - \partial_x\left( e^{d-a+2c} \Phi_x \right) &=& 0,
\label{Eq:Ev3}
\end{eqnarray}
which are subject to the constraints
\begin{eqnarray}
{\cal H} &:=& e^{d-a}\left[ 2c_{xx} + (3c_x - 2a_x)c_x \right] 
 - e^{a-d} c_t(2a_t + c_t) - e^{a+d-2c}
 + \frac{\kappa_0}{2}\, Q^2 e^{a+d-4c}
 + \frac{\kappa}{2}\left[ e^{a-d}\Phi_t^2 + e^{d-a} \Phi_x^2 \right] = 0,
 \label{Eq:HamConstraint}\\
{\cal M} &:=& 2c_{tx} + 2c_t c_x - 2d_x c_t - 2a_t c_x 
 + \kappa\, \Phi_t \Phi_x = 0.
\label{Eq:MomConstraint}
\end{eqnarray}
Here, the subscript $t$ and $x$ refer to the derivatives with respect to $t$ and $x$, respectively.

For a static configuration, the scalar field $\Phi$ and the metric coefficients $d$, $a$ and $c$ are independent of $t$. In this case the field equations can be integrated analytically. For zero charge the corresponding solutions have been obtained in \cite{hE73,kB73}. Here, we generalize their solutions for the charged case. In the static case, the field equations simplify to
\begin{eqnarray}
\left[ e^{d-a+2c} d_x \right]_x &=& \frac{\kappa_0}{2} Q^2\, e^{a+d-2c},
\label{Eq:Static1}\\
\left[ e^{d-a+2c} c_x \right]_x &=& e^{a+d} 
 - \frac{\kappa_0}{2} Q^2\, e^{a+d-2c},
\label{Eq:Static2}\\
\left[ e^{d-a+2c}\Phi_x \right]_x &=& 0,
\label{Eq:Static3}\\
(2d_x + c_x)c_x &=& e^{2(a-c)} - \frac{\kappa_0}{2} Q^2\, e^{2(a-2c)} 
 + \frac{\kappa}{2}\Phi_x^2\, .
\label{Eq:Static4}
\end{eqnarray}
Adopting a gauge where $a=-d$, the first two equations imply that $[e^{2(c+d)}]_{xx} = 2$ which has the general solution $e^{2(c+d)} = x^2 + 2\alpha_1 x + \alpha_0$ with two constants $\alpha_0$ and $\alpha_1$. By a suitable translation of the coordinate $x$ it is always possible to obtain $\alpha_1=0$. Furthermore, since we are interested in wormhole geometries with the properties described below equation~(\ref{Eq:SphericalMetric}) we need $e^{2(c+d)} > 0$ for all $x\in\Real$. Therefore, we have $e^{2(c+d)} = x^2 + b^2$ with some strictly positive constant $b > 0$. Equation (\ref{Eq:Static3}) then gives $\Phi = \Phi_1\arctan(x/b) + \Phi_0$ with two integration constants $\Phi_0$ and $\Phi_1$. Since only the gradient of $\Phi$ appears in the equations we set the parameter $\Phi_0$ to zero in what follows. Next, setting $\epsilon :=
\sqrt{\kappa_0 Q^2/2b^2}$ and $y := \arctan(x/b)\in (-\pi/2,\pi/2)$,
Eq. (\ref{Eq:Static1}) gives
\begin{equation}
d_{yy} = \epsilon^2 e^{2d}.
\end{equation}
The unique local solution with initial conditions $\left. d
\right|_{y=0} = \gamma_0\in\Real$ and $\left. d_y \right|_{y=0} =
\gamma_1\in\Real$ is
\begin{equation}
d = \gamma_0 
  - \log\left[ \cosh(\Lambda y) - \gamma_1\frac{\sinh(\Lambda y)}{\Lambda}
        \right],\qquad
\Lambda := \sqrt{\gamma_1^2 - e^{2\gamma_0}\epsilon^2}\, .
\end{equation}
We distinguish between the subcritical case where $\Lambda > 0$, the
critical case where $\Lambda=0$ and $d = \gamma_0 - \log(1-\gamma_1
y)$ and the supercritical case where $\Lambda = i\mu$ for some real,
strictly positive number $\mu$ in which case $d = \gamma_0 -
\log\left[ \cos(\mu y) - \gamma_1\frac{\sin(\mu y)}{\mu} \right]$. In
order to obtain a global wormhole solution, we need the expression
inside the square brackets to be strictly positive for all $y\in
[-\pi/2,\pi/2]$. This is the case if and only if
\begin{equation}
\left\{
\begin{array}{rl}
  \frac{\tanh\left(\Lambda\frac{\pi}{2}\right)}{\Lambda} |\gamma_1| < 1
& \hbox{in the subcritical case},\\
  \frac{\pi}{2}|\gamma_1| < 1
& \hbox{in the critical case},\\
  \mu < 1 \hbox{ and }
  \frac{\tan\left(\mu\frac{\pi}{2}\right)}{\mu} |\gamma_1| < 1
& \hbox{in the supercritical case}
\end{array} \right\}.
\label{Eq:WormholeParamConstr1}
\end{equation}
Finally, Eq. (\ref{Eq:Static4}) yields the relation
\begin{equation}
-\kappa\Phi_1^2 = 2(1 + \Lambda^2)
\label{Eq:WormholeParamConstr2}
\end{equation}
between the parameters $\Lambda$ and $\Phi_1$.

Summarizing, we obtain the solutions
\begin{eqnarray}
\Phi &=& \Phi_1 y,
\label{Eq:WormholeSF}\\
F &=& \frac{Q_e}{b} e^{2d} dt\wedge dy 
   + Q_m\,  d\vartheta \wedge \sin\vartheta\; d\varphi,
\label{Eq:WormholeEMF}\\
ds^2 &=& -e^{2d} dt^2 + e^{-2d}\left[
 dx^2 + (x^2 + b^2)\left( d\vartheta^2 + \sin^2\vartheta\; d\varphi^2 \right)
 \right],
\label{Eq:WormholeMetric}
\end{eqnarray}
where $y = \arctan(x/b)$ and $e^{2d} = e^{2\gamma_0}\left[ \cosh(\Lambda
  y) - \gamma_1\frac{\sinh(\Lambda y)}{\Lambda} \right]^{-2}$ with
$\Lambda = \sqrt{\gamma_1^2 - \kappa_0(Q_e^2 + Q_m^2)
  e^{2\gamma_0}/(2b^2)}$. The parameters $b$, $\Phi_1$, $Q_e$, $Q_m$,
$\gamma_0$ and $\gamma_1$ are subject to the two constraints
(\ref{Eq:WormholeParamConstr1}) and
(\ref{Eq:WormholeParamConstr2}). Notice that the constant rescaling $t
\mapsto \exp(-\Omega) t$, $x\mapsto \exp(\Omega) x$, $b \mapsto
\exp(\Omega) b$, $\gamma_0 \mapsto \gamma_0 + \Omega$, $\gamma_1 \to
\gamma_1$, $Q_e \mapsto Q_e$, $Q_m \mapsto Q_m$ with $\Omega$ a
nonvanishing constant leaves the solution unchanged. In particular, we
can rescale the coordinates such that either $\lim\limits_{x\to+\infty}
d = 0$ or $\lim\limits_{x\to-\infty} d = 0$ which shows that the
spacetime described by the metric (\ref{Eq:WormholeMetric}) has indeed
two asymptotically flat ends at $x\to +\infty$ and $x\to -\infty$,
respectively.

Therefore, we obtain a four-parameter family of wormhole solutions characterized by the scale invariant quantities $B := b e^{-\gamma_0} > 0$, $\gamma_1\in\Real$, $Q_e\in\Real$ and $Q_m\in\Real$, which are subject to the restrictions (\ref{Eq:WormholeParamConstr1}). In the particular case $Q_e = Q_m = 0$ this family reduces to the bi-parametric solution obtained in \cite{hE73,kB73}. As shown in \cite{jGfGoS09a,jGfGoS09b} these uncharged wormholes are unstable with respect to linear and nonlinear perturbations.

Let us analyze the physical properties of the wormhole solutions. First, the wormhole throat is given by the global minimum of the areal radius $r = \sqrt{x^2 + b^2} e^{-d}$. In Lemma \ref{Lem:Area} in the appendix we show that $r$ has a unique minimum which is determined by the unique root of the function $c_x$ given in Eq. (\ref{Eq:cx}) below. Next, we compute the Misner-Sharp mass function \cite{cMdS64}. For the spherically symmetric spacetime metric given by equation~(\ref{Eq:SphericalMetric}) it is defined by
\begin{displaymath}
m(t,x) := \frac{r}{2}\left[ 1 - \tilde{g}(dr,dr) \right] 
 = \frac{e^c}{2}\left[ 1 + e^{2(c-d)} c_t^2 - e^{2(c-a)} c_x^2 \right].
\end{displaymath}
Specialized to the static family of solutions described by
Eqs.~(\ref{Eq:WormholeSF},\ref{Eq:WormholeEMF},\ref{Eq:WormholeMetric})
it yields
\begin{displaymath}
m(x) = \frac{r}{2}\left[ 1 - (x^2 + b^2) c_x^2 \right],
\end{displaymath}
where $c_x$ is the derivative of the logarithm of the areal radius,
\begin{equation}
c_x = \frac{b}{x^2 + b^2}\left[ \tan(y) - \Lambda
 \frac{\gamma_1\cosh(\Lambda y) - \Lambda\sinh(\Lambda y)}
      {\Lambda\cosh(\Lambda y) - \gamma_1\sinh(\Lambda y)} \right].
\label{Eq:cx}
\end{equation}
The ADM masses of the two asymptotically flat ends can be computed by
considering the asymptotic values $m_{\pm\infty} :=
\lim\limits_{x\to\pm\infty} m(x)$ of the mass function, which yields
\begin{eqnarray}
m_{+\infty} &=& B\left[ \gamma_1\cosh\left(\Lambda\frac{\pi}{2}\right) 
 - \Lambda\sinh\left(\Lambda\frac{\pi}{2}\right) \right],\\
m_{-\infty} &=& -B\left[ \gamma_1\cosh\left(\Lambda\frac{\pi}{2}\right) 
 + \Lambda\sinh\left(\Lambda\frac{\pi}{2}\right) \right].
\end{eqnarray}
In particular, we have the relations
\begin{eqnarray}
m_{+\infty} - m_{-\infty} &=& 2B\gamma_1\cosh\left(\Lambda\frac{\pi}{2}\right),
\nonumber\\
m_{+\infty} + m_{-\infty} &=& -2B\Lambda\sinh\left(\Lambda\frac{\pi}{2}\right),
\nonumber\\
m_{+\infty} m_{-\infty} &=& -B^2\left[ \gamma_1^2 
 + \nu^2\sinh^2\left(\Lambda\frac{\pi}{2}\right) \right],
\nonumber
\end{eqnarray}
where we have set $\nu := \sqrt{\kappa_0(Q_e^2 + Q_m^2)/(2B^2)}$. From the first relation and Eqs.~(\ref{Eq:WormholeSF},\ref{Eq:WormholeEMF},\ref{Eq:WormholeMetric}) we see that the wormholes are reflection-symmetric about their throat if and only if $\gamma_1=0$. Since $\Lambda = \sqrt{\gamma_1^2 - \nu^2}$ the asymmetry parameter $\gamma_1$ and the dimensionless charge $\nu$ determine the asymptotic masses $m_{+\infty}$ and $m_{-\infty}$, the total electromagnetic charge $Q := \sqrt{Q_e^2 + Q_m^2}$ and the areal radius of the throat up to the scale factor $B$. From the second and third relations we see that in the subcritical case ($\Lambda > 0$)  the two masses have opposite signs and that their sum is negative. In the critical case ($\Lambda=0$) the sum of the two masses is zero, and the masses are different from zero unless $\gamma_1 = Q_e = Q_m = 0$ in which case $m_{+\infty} = m_{-\infty} = 0$. In the supercritical case ($\Lambda = i\mu$, $\mu > 0$), the sum of the two masses is positive, with opposite signs if $\gamma_1^2 > \mu^2\tan^2(\mu\pi/2)$ and equal signs if $\gamma_1^2 < \mu^2\tan^2(\mu\pi/2)$ while one of the masses is zero and the other is positive if $\gamma_1 = \pm \mu\tan(\mu\pi/2)$.

\section{Derivation of the pulsation and master equations}
\label{Sec:Pulsation}

In this section we derive the relevant equations for analyzing the linear stability of the four-parameter family of static wormhole solutions discussed in the previous section. For this, we consider small perturbations of the form
\begin{displaymath}
\Phi(\lambda) = \Phi + \lambda\delta\Phi + {\cal O}(\lambda^2),
\end{displaymath}
where $\Phi$ is the background solution, and where
\begin{displaymath}
\delta\Phi := \left. \frac{d}{d\lambda} \Phi(\lambda) \right|_{\lambda=0}
\end{displaymath}
denotes the variation of $\Phi$. The same applies to the other fields
$d$, $a$ and $c$. A general method for analyzing such perturbations has been 
developed in \cite{oBmHnS96}. Since in spherical symmetry there are no gravitational nor electromagnetic dynamical degrees of freedom, one obtains a single master equation for the linearized scalar field $\delta\Phi$. However, as we have discussed in detail in \cite{jGfGoS09a} for the uncharged case, the resulting master equation turns out to be singular at the throat which leads to difficulties when studying the stability by standard methods based on Schr\"odinger operators. For this reason, we will not use the method described in \cite{oBmHnS96} and instead base our treatment on a gauge-invariant approach which leads to a constrained wave system which is everywhere regular.

\subsection{Gauge-invariant quantities}

With respect to an infinitesimal coordinate transformation $\delta t \mapsto \delta t + \xi^t$, $\delta x \mapsto \delta x + \xi^x$ on the $2$-manifold $\tilde{M}$ generated
by a vector field $(\xi^t,\xi^x)$, we have
\begin{equation}
\delta a \mapsto \delta a + e^{-a} (e^a\xi^x)_x\; ,\qquad
\delta c \mapsto \delta c + \xi^x c_x\; ,\qquad
\delta\Phi \mapsto \delta\Phi + \xi^x\Phi_x\; .
\end{equation}
Since $\Phi_x\neq 0$ everywhere we may construct the following two gauge-invariant fields,
\begin{equation}
A:=\delta a - e^{-a}\left( e^a\frac{\delta\Phi}{\Phi_x} \right)_x\; , \qquad
C:=\delta c - c_x\frac{\delta\Phi}{\Phi_x}\; ,
\end{equation}
which reduce to $\delta a$ and $\delta c$, respectively, in the gauge
$\delta\Phi=0$.

\subsection{Constrained wave system for $A$ and $C$}

Here we derive a constrained wave system for the gauge-invariant quantities $A$ and $C$. In order to do so, we first rescale the coordinate $x$ such that $b=1$. In terms of the coordinate $y = \arctan(x)$ which satisfies $\partial_y = e^{d-a+2c}\partial_x$, the background Eqs. (\ref{Eq:Static1}--\ref{Eq:Static3}) yield
\begin{equation}
d_{yy} = \epsilon^2 e^{2d}, \qquad
c_{yy} = e^{2(d+c)} - \epsilon^2 e^{2d}, \qquad
\Phi_{yy} = 0,
\label{Eq:Background}
\end{equation}
where $\epsilon^2 = \kappa_0 Q^2/2$.

Next, we consider the constraint equations (\ref{Eq:HamConstraint},\ref{Eq:MomConstraint}). Their linearization yields
\begin{eqnarray}
\delta( e^{a-d} {\cal H}) &=& 2\delta c_{xx} + (6c_x - 2a_x)\delta c_x 
 - 2c_x\delta a_x - 2(\delta a - \delta c)e^{2(a-c)} 
 + \kappa_0 Q^2(\delta a - 2\delta c) e^{2a-4c} + \kappa\Phi_x\delta\Phi_x
 = 0,\\
\delta {\cal M} &=& 2\delta c_{tx} + 2(c_x - d_x)\delta c_t - 2c_x\delta a_t 
 + \kappa\, \Phi_x\delta\Phi_t = 0.
\end{eqnarray}
With the help of the background equations (\ref{Eq:Background}) one
may rewrite this as
\begin{eqnarray}
\frac{1}{2}\, e^{2(d-a+2c)}\delta( e^{a-d} {\cal H})
 &=& \left[ \delta c_y + (c_y-d_y)\delta c - c_y\delta a 
 + \frac{\kappa}{2}\Phi_y \delta\Phi \right]_y = 0, \\
\frac{1}{2}\, e^{d-a+2c}\delta{\cal M} 
 &=& \left[ \delta c_y + (c_y-d_y)\delta c - c_y\delta a 
 + \frac{\kappa}{2}\Phi_y \delta\Phi \right]_t = 0,
\end{eqnarray}
which shows that the expression inside the square bracket must be
equal to a constant $\sigma$. In terms of the gauge-invariant
quantities $A$ and $C$ defined above, the resulting first integral is
\begin{equation}
C_y + (c_y - d_y)C - c_y A = \sigma.
\label{Eq:FirstIntegral}
\end{equation}
The interpretation of the constant $\sigma$ is the following. With
respect to an infinitesimal variation of the constants $B\gamma_1$ and
$\Lambda$ (keeping the charges $Q_e$ and $Q_m$ fixed), the family of
static solutions (\ref{Eq:WormholeMetric}) yields the linearized
solution
\begin{equation}
C = \frac{d_y}{\Lambda^2} \delta(B\gamma_1) 
  - F\frac{\delta\Lambda}{\Lambda}\; ,\qquad
A = -(1 + xy)\frac{\Lambda\delta\Lambda}{1+\Lambda^2} + C,
\label{Eq:LinearizedStaticSolution}
\end{equation}
where the function $F$ is given by
\begin{equation}
F = 1 + xy - \frac{y c_y}{1+\Lambda^2}
  = 1 + \frac{(\Lambda^2 x + d_y)y}{1+\Lambda^2}\, .
\label{Eq:DefF}
\end{equation}
Introducing the expressions (\ref{Eq:LinearizedStaticSolution}) in (\ref{Eq:FirstIntegral}) gives $\sigma = -\delta(B\gamma_1)$. Therefore, $\sigma$ describes variations of the static family of wormhole solutions with respect to the constant $B\gamma_1$. Since any solution to the linearized equations may be written as the sum of such a variation plus a solution with $\delta(B\gamma_1) = 0$ we may assume that $\sigma=0$ in the following.

Next, we linearize the evolution Eqs.
(\ref{Eq:Ev1},\ref{Eq:Ev2},\ref{Eq:Ev3}). For simplicity, we choose
the gauge such that $\delta\Phi = 0$, in which case $A = \delta a$ and
$C = \delta c$. Linearization of Eq. (\ref{Eq:Ev3}) yields $\delta d -
\delta a + 2\delta c = h(t)$ for some function $h(t)$ which we may set
to zero by a redefinition of $\delta t$. Using $\delta d = \delta a -
2\delta c$ and the first integral (\ref{Eq:FirstIntegral}) with
$\sigma=0$ in the linearization of the evolution Eqs.
(\ref{Eq:Ev1},\ref{Eq:Ev2}) a lengthy calculation yields the
constrained wave system
\begin{equation}
u_{tt} - e^{-2c}\left[ e^{-2c} u_y \right]_y + V u = 0,\qquad
{\cal C} := (u_2)_y + (c_y-d_y) u_2 - c_y u_1 = 0, 
\label{Eq:ConstrainedPulsation}
\end{equation}
where we have defined
\begin{eqnarray}
u = e^{-c}\left( \begin{array}{r} A - C \\ C \end{array} \right),\qquad
V = e^{-4c}\left( \begin{array}{ll}
  3 c_y^2 + 4c_y d_y - 3e^{2(d+c)} + 5\epsilon^2 e^{2d} 
& 4\Lambda^2 \\
  -4c_y^2 + 2 e^{2(d+c)} - 2\epsilon^2 e^{2d} 
& 3c_y^2 - 4c_y d_y - e^{2(d+c)} + 3\epsilon^2 e^{2d}
 \end{array} \right).
\end{eqnarray}
The constrained wave system (\ref{Eq:ConstrainedPulsation}) describes the dynamics of the two gauge-invariant linearized fields $A$ and $C$. The linear stability properties of the wormholes are determined by the Cauchy evolution of this system. Two difficulties with analyzing the properties of the solutions are the fact that we are confronted with a coupled system of two equations (as opposed to a single, scalar equation) and the presence of the constraint ${\cal C} = 0$. In the next subsection we start by deriving a decoupled equation for the constraint field ${\cal C}$ based on a factorization method. This shows that it is sufficient to enforce the constraint and its time-derivative at an initial time. As a byproduct of our factorization method, we also obtain a master equation for a
quantity $v_1$ defined below, from which $u_1$ and $u_2$ can be reconstructed. However, this equation turns out to be singular at the throat, and as mentioned before, this means one has to be careful with the stability analysis. For this reason, we derive in the following subsection a different master equation which, in the critical and subcritical cases is everywhere regular and allows to prove that such wormholes are linearly
unstable. In the supercritical case, however, both master equations turn out to be singular, and so the constrained wave system (\ref{Eq:ConstrainedPulsation}) has to be analyzed directly. If the potential $V$ would be symmetric, or if it could be brought into symmetric form by a linear transformation of $u$, one could analyze the system by spectral analysis of the formally self-adjoint operator ${\cal H} = -e^{-2c}\partial_y e^{-2c}\partial_y + V$. The transformation $w_1 = u_1$, $w_2 = -u_1 + u_2$ brings the system into the form
\begin{displaymath}
w_{tt} - e^{-2c}\left[ e^{-2c} w_y \right]_y +  \bar{V} w = 0,\qquad
\bar{V} = e^{-4c}\left( \begin{array}{rr} 
 D + E & 4\Lambda^2 \\
 4 & D - E
\end{array} \right),
\end{displaymath}
with $D = 3c_y^2 - 2e^{2(c+d)} + 4\epsilon^2 e^{2d}$ and $E = 4c_y d_y - e^{2(c+d)} + \epsilon^2 e^{2d} + 4\Lambda^2$. If $\Lambda^2 > 0$ this can be symmetrized by a trivial rescaling of $w$; for $\Lambda^2 < 0$, however, $\bar{V}_{12}$ and $\bar{V}_{21}$ have different signs. In fact, we will show in the next section that in the latter case the operator ${\cal H}$ may have complex eigenvalues, and so it cannot be written as a symmetric operator.

\subsection{Factorization of the Hamilton operator and master equation I}

Consider the two-channel Schr\"odinger operator
\begin{displaymath}
{\cal H} := -\partial^2 + V, \qquad
\partial := e^{-2c}\partial_y = e^{d-a}\partial_x\; .
\end{displaymath}
Here, we try to factorize it in the form ${\cal H} = {\cal A}{\cal
  B}$, with the two first-order operators
\begin{displaymath}
{\cal A} = \partial + K, \qquad
{\cal B} = -\partial + K,
\end{displaymath}
where $K$ is a $2\times 2$ matrix which has to satisfy the Riccati matrix equation
\begin{equation}
\partial K + K^2 = V.
\label{Eq:Riccati}
\end{equation}
If $K$ solves (\ref{Eq:Riccati}), the factorization ${\cal H} = {\cal
  A}{\cal B}$ allows us to rewrite the wave problem $u_{tt} + {\cal
  H} u = 0$ into first-order form,
\begin{eqnarray}
u_t &=& {\cal A} v,
\label{Eq:PulsationFirstOrder1}\\
v_t &=& -{\cal B} u.
\label{Eq:PulsationFirstOrder2}
\end{eqnarray}
In particular, it follows that $v$ satisfies the dual wave problem
\begin{equation}
v_{tt} + {\cal B}{\cal A} v = 0,
\label{Eq:TransformedPulsation}
\end{equation}
where ${\cal B}{\cal A} = -\partial^2 + \hat{V}$ with the transformed
potential $\hat{V} = -V + 2K^2$.

A solution to (\ref{Eq:Riccati}) can be obtained as follows. We demand that the second component of $v_t = -{\cal B}u$ is proportional to the constraint variable ${\cal C}$ defined in Eq. (\ref{Eq:ConstrainedPulsation}). This implies that $K$ must be of the form
\begin{equation}
K = e^{-2c}\left( \begin{array}{ll}
f & g \\ c_y & d_y - c_y \end{array} \right)
\end{equation}
with two unkown functions $f$ and $g$. (Notice that in this case
$\partial_t v_2 = e^{-2c}{\cal C}$.) Introducing this ansatz into
Eq. (\ref{Eq:Riccati}) yields the unique solution
\begin{displaymath}
f = d_y + \frac{1+\Lambda^2}{c_y}\; ,\qquad
g = -\frac{\Lambda^2}{c_y}\; ,
\end{displaymath}
for $f$ and $g$. The transformed potential is then
\begin{equation}
\hat{V} = e^{-4c}\left( \begin{array}{ll}
  c_{yy}\left[ 1 + \frac{2(1+\Lambda^2)}{c_y^2} \right] - c_y^2 & 
 -\frac{2\Lambda^2}{c_y^2} c_{yy} \\
 0 & c_{yy} - c_y^2
 \end{array} \right).
\end{equation}
Therefore, the constraint variable $v_2$ satisfies a decoupled wave
equation with potential $\hat{V}_{22} = e^{-4c}(c_{yy} - c_y^2)$, and
it is consistent to enforce the constraint ${\cal C}=0$. Setting
$v_2=0$ the first-order system
(\ref{Eq:PulsationFirstOrder1},\ref{Eq:PulsationFirstOrder2}) then
reduces to
\begin{eqnarray}
\partial_t u_1 &=& \partial v_1 + e^{-2c} f v_1\; ,
\label{Eq:u1t}\\
\partial_t u_2 &=& e^{-2c} c_y v_1\; ,
\label{Eq:u2t}\\
\partial_t v_1 &=& \partial u_1 - e^{-2c}(f u_1 + g u_2).
\label{Eq:v1t}
\end{eqnarray}
In particular, $v_1$ satisfies the following master equation,
\begin{equation}
\left[ \partial_t^2 - \partial^2 + \hat{V}_{11} \right] v_1 = 0, \qquad
\hat{V}_{11} = e^{-4c}\left(
c_{yy}\left[ 1 + \frac{2(1+\Lambda^2)}{c_y^2} \right] - c_y^2 \right).
\label{Eq:MasterI}
\end{equation}
As mentioned above, the resulting potential $\hat{V}_{11}$ is singular at the wormhole throat, where $c_y=0$. As discussed in detail in \cite{jGfGoS09a} this enforces an unphysical boundary condition at the throat if one tries to define $-\partial^2 + \hat{V}_{11}$ as a self-adjoint operator. Namely, it requires $v_1$ to approach zero sufficiently rapidly as $y$ converges to the throat's location. However, there is no reason for enforcing such a strong condition on $v_1$ from a physical point of view. As we will see in the next section, physically permissible perturbations even allow $v_1$ to diverge at the throat. In particular, this implies that the operator $-\partial^2 + \hat{V}_{11}$ is not symmetric when defined on the space of physically permissible states.

As we show next, it is possible to obtain a different master equation which in the subcritical and critical cases is everywhere regular and yields a self-adjoint operator without enforcing unphysical boundary conditions.

\subsection{Master equation II}

The derivation of the new master equation is based on the observation that the constrained wave system (\ref{Eq:ConstrainedPulsation}) possesses the particular solution
\begin{equation}
u^{static} = \left( \begin{array}{r} G \\ H \end{array} \right)
           = e^{-c}\left( \begin{array}{r} 
 \frac{\Lambda^2}{1+\Lambda^2}(1+xy) \\ F \end{array} \right)
 \label{Eq:ExactStaticMode}
\end{equation}
which is obtained from (\ref{Eq:LinearizedStaticSolution}) after
setting $\delta(B\gamma_1)=0$ and $\delta\Lambda = -\Lambda$. A
related solution is obtained by multiplication with $t$, corresponding
to the following solution of the first-order system (\ref{Eq:u1t},\ref{Eq:u2t},\ref{Eq:v1t})
\begin{displaymath}
u = t u^{static},\qquad
v_1 = \Psi_0 \equiv \frac{e^c F}{c_y}\; .
\end{displaymath}
Since $\Psi_0$ is a time-independent solution of the master equation
(\ref{Eq:MasterI}) we may rewrite the latter in first-order form
\begin{eqnarray}
\partial_t v_1 &=& \left( \partial + \frac{\partial\Psi_0}{\Psi_0} \right)\chi,\\
\partial_t\chi &=& \left( \partial - \frac{\partial\Psi_0}{\Psi_0} \right) v_1\; .
\end{eqnarray}
The corresponding dual wave equation is
\begin{equation}
\left[ \partial_t^2 - \partial^2 + W \right]\chi = 0, \qquad
\label{Eq:MasterII}
\end{equation}
with the transformed potential
\begin{eqnarray}
W &=& -\hat{V}_{11} + 2\left( \frac{\partial\Psi_0}{\Psi_0} \right)^2
\nonumber\\
  &=& e^{-4c}\left\{ -3(1+\Lambda^2) + 2d_y(d_y-c_y)
 - 4\frac{G}{H}\left[ c_y d_y + 1 + \Lambda^2  \right] 
 + 2c_y^2\left( \frac{G}{H} \right)^2 \right\},
\end{eqnarray}
where we have used the identity $\frac{\partial_y\Psi_0}{\Psi_0} + f = \frac{G}{H} c_y$.
In terms of the new variable $\chi$ the first-order system yields
\begin{eqnarray}
\partial_t u_1 &=& \partial_t\chi + e^{-2c}\frac{G}{H}(c_y v_1),
\label{Eq:NewFirstOrder1}\\
\partial_t u_2 &=& e^{-2c}(c_y v_1),
\label{Eq:NewFirstOrder2}\\
\partial_t(c_y v_1) &=& c_y\partial\chi 
 + e^{-2c}\left (\frac{G}{H} c_y^2 - c_y d_y - 1 - \Lambda^2 \right)\chi,\label{Eq:NewFirstOrder3}
\end{eqnarray}
which allows to obtain the gauge-invariant perturbation quantities $u_1$, $u_2$ (and $c_y v_1$) from $\chi$ after a time integration.

In contrast to the first master equation, the new master equation
(\ref{Eq:MasterII}) is regular at the throat. In fact, the potential
$W$ is everywhere regular as long as the function $F$ does not have
any zeroes. This turns out to be the case for the critical and
subcritical cases, see Lemma \ref{Lem:F} in the appendix.

\section{Linear stability analysis}
\label{Sec:LinStabAnal}

In this section we discuss the linear stability of the wormhole solutions in the subcritical, critical and supercritical cases. In the first two cases, we show instability by proving that the master equation (\ref{Eq:MasterII}) which is regular in those cases possesses precisely one exponentially in time growing mode. In the supercritical case, both master equations are singular, and we analyze the stability by different means.

\subsection{The subcritical case}

The results from the previous section allow us to describe the stability problem in the subcritical case by the regular master equation (\ref{Eq:MasterII}) on the Hilbert space $X = L^2(\Real, e^{-2d} dx)$ which admits the zero mode
\begin{equation}
\chi_0 = \frac{1}{\Psi_0} = \frac{c_y}{e^c F}\, .
\end{equation}
Since the function $c_y/e^c$ is uniformly bounded on $-\infty < x < +\infty$ and possesses exactly one zero, and since the function $F$ is strictly positive and satisfies $F/x \to \Lambda^2(1 + \Lambda^2)^{-1} \pi/2$ for $x\to\pm\infty$, this mode belongs to $X$ and represents a bound state of the Schr\"odinger operator $-\partial^2 + W$. Because it has one node, it follows from the nodal theorem\footnote{See, for instance, \cite{CourantHilbert-Book} or \cite{hApQ95} for a generalization to systems.} 
that it is the first excited bound state and so the operator $-\partial^2 + W$ possesses precisely one negative eigenvalue $-\beta^2 < 0$ with eigenfunction $\chi_\beta$, corresponding to an exponentially growing mode of (\ref{Eq:MasterII}) which is of the form
\begin{equation}
\chi(t,x) = e^{\beta t}\chi_\beta(x).
\end{equation}
Since the coefficients in the Eqs. (\ref{Eq:NewFirstOrder1},\ref{Eq:NewFirstOrder2},\ref{Eq:NewFirstOrder3}) are everywhere regular this gives rise to a unique unstable mode for
each subcritical wormhole. We conclude that all such wormholes are unstable with respect to linear, spherically symmetric perturbations.

\subsection{The critical case}

Linear perturbations in the critical case are also described by the regular master equation (\ref{Eq:MasterII}). However, in contrast to the previous case, the function
\begin{displaymath}
\chi_0 = 1/\Psi_0 
 = \frac{ x(1-\gamma_1 y) - \gamma_1 }{\sqrt{1+x^2} (1 - \gamma_1 y)^2}
 \end{displaymath}
is not normalizable, so the above argument based on the nodal theorem does not directly apply. Instead, we construct a family $f_n$, $n=1,2,3,...$, of static solutions to (\ref{Eq:MasterII}) which are defined on the interval $-n \leq x < +\infty$ and satisfy the following two properties for large enough $n$: (i) $f_n(-n) = 0$, (ii) $(f_n)_x(-n) \neq 0$, (iii) $f_n$ has exactly one zero on the interval $x > n$. It then follows from the results in Ref. \cite{hApQ95} that there is a unique bound state with negative energy, as in the subcritical case. Therefore, all critical wormholes are linearly unstable as well.

The family $f_n$ is defined as follows. Let $x_{throat}$ be the value of $x$ at the throat, where $c_y=0$, and let $n > -x_{throat}$. Then,
\begin{displaymath}
f_n(x) := \chi_0(x)\cdot\left\{ \begin{array}{ll}
 \int\limits_{-n}^x \frac{e^{-2d(\bar{x})} d\bar{x}}{\chi_0(\bar{x})^2}  
 &, -n \leq x < x_{throat} , \\
 k - \int\limits_x^n \frac{e^{-2d(\bar{x})} d\bar{x}}{\chi_0(\bar{x})^2}  
 &, x > x_{throat} ,
\end{array} \right.
\end{displaymath}
where $k$ is a constant to be determined. It is simple to verify that $f_n$ satisfies
the relation $\chi_0\partial f_n = 1 + f_n\partial\chi_0$ and the master equation (\ref{Eq:MasterII}) on the two open intervals $(-n,x_{throat})$, $(x_{throat},+\infty)$, and that $f_n(-n)=0$ and $(f_n)_x(-n) < 0$. In order to analyze the behavior of $f_n$ near $x=x_{throat}$ we first introduce the new coordinate
\begin{equation}
\rho = R(x) := \int\limits_{x_{throat}}^x e^{-2d(\bar{x})} d\bar{x},\qquad
-\infty < x < +\infty,
\label{Eq:rhoDef}
\end{equation}
in terms of which we have $\partial = \partial_\rho$. Next, since $\chi_0$ satisfies $(-\partial^2 + W)\chi_0 = 0$ and vanishes at $\rho=0$, we can write it in the form
$\chi_0(\rho) = \alpha\rho[ 1 + \rho^2 q(\rho) ]$, where $q$ is a smooth function on $\Real$ such that $q(0)\neq 0$, and $\alpha > 0$. Also, $1 + \rho^2 q(\rho)$ has to be strictly positive since otherwise $\chi_0$ would have more than one zero. In terms of this, we find
\begin{displaymath}
f_n(\rho) = -\frac{1 + \rho^2 q(\rho)}{\alpha}\cdot\left\{ \begin{array}{ll}
 1 - \frac{\rho}{R(-n)} + \rho\int\limits_{R(-n)}^\rho \frac{ 2q(\bar{\rho}) + \bar{\rho}^2 q(\bar{\rho})^2 }{[ 1 + \bar{\rho}^2 q(\bar{\rho}) ]^2}  d\bar{\rho}
 &, R(-n) \leq \rho < 0, \\
 1 - \frac{\rho}{R(n)} - \rho\int\limits_{\rho}^{R(n)} \frac{ 2q(\bar{\rho}) + \bar{\rho}^2 q(\bar{\rho})^2 }{[ 1 + \bar{\rho}^2 q(\bar{\rho}) ]^2} d\bar{\rho} - \alpha^2 k\rho 
 &, \rho > 0 ,
\end{array} \right.
\end{displaymath}
which shows that $f_n$ is continuous at $\rho=0$. Furthermore, we see that $(f_n)_\rho$ is also continuous at $\rho=0$ provided we choose $k$ such that
\begin{displaymath}
\alpha^2 k = \frac{1}{R(-n)} - \frac{1}{R(n)} - \int\limits_{R(-n)}^{R(n)}\frac{ 2q(\bar{\rho}) + \bar{\rho}^2 q(\bar{\rho})^2 }{[ 1 + \bar{\rho}^2 q(\bar{\rho}) ]^2} d\bar{\rho}\, .
\end{displaymath}
Therefore, $f_n$ can be extended on the whole interval $-n < x < +\infty$. Finally, we notice that $f_n$ is negative on the interval $-n < x \leq x_{throat}$ and positive for large enough $x > x_{throat}$. Hence, $f_n$ has at least one zero at some $x^* > x_{throat}$. On the other hand, because $\chi_0(x^*)\partial f_n(x^*) = 1 + f_n(x^*)\partial\chi_0(x^*) = 1$ and $\chi_0(x^*) > 0$ it follows that $\partial f_n(x^*) > 0$ which means that this zero is unique. In the reflection-symmetric case where $\gamma_1 = 0$ we have $\chi_0(x) = x/\sqrt{1+x^2}$ and $f_n(x) = (x+n)(x-1/n)/\sqrt{1+x^2}$.

\subsection{The supercritical case: Numerical integration of the constraint wave problem}

For the supercritical case, $F$ has always two zeroes and both master
equations are singular. Therefore, the previous arguments cannot be
used to establish the linear instability of the wormholes in this case. Short of an analytic proof, we shall analyze the stability of supercritical wormholes by numerical means. We start in this subsection with a numerical integration scheme of the constraint wave problem (\ref{Eq:ConstrainedPulsation}). We model this scheme on the recent work in Ref. \cite{aZdNsH09}, where hyperboloidal
time slices with a compactified space coordinate are used. One of the
main advantages of this scheme is that the Cauchy evolution is
performed on a compactified domain where the boundaries of the domain
correspond to future null infinity. In this way, one avoids the
problem of introducing an artificial timelike boundary with absorbing
boundary conditions.  This is particularly attractive for our
stability problem since we do not want the time evolution of our
solution to be contaminated by artificial boundary conditions.

The starting point of the numerical scheme is to rewrite the wave
equation in Eq. (\ref{Eq:ConstrainedPulsation}) in geometric form
\begin{equation}
-\tilde{g}^{ab}\tilde{\nabla}_a\tilde{\nabla}_b u + \tilde{V} u = 0,
\label{Eq:GeometricWaveEq}
\end{equation}
where $\tilde{g} = -e^{2d} dt^2 + e^{-2d} dx^2$ is the wormhole
$2$-metric, $\tilde{\nabla}$ is the associated connection, and
$\tilde{V} = e^{-2d} V$ is the rescaled potential. An observation of
later importance is the fact that the rescaled potential decays at
least as ${\cal O}(x^{-2})$ when $x\to\pm\infty$.

Next, we introduce a new time coordinate $\tau := t - h(x)$, where the
height function $h$, which will be specified later, is such that the
$\tau=const$ slices are everywhere spacelike and asymptote to outgoing
null geodesics as $x\to\pm\infty$. This means that the function $e^{4d}
h_x^2$ is strictly less than one and converges to one as
$x\to\pm\infty$. The $\tau=const$ slices, as embedded in the
four-dimensional, spherically symmetric spacetime with $2$-metric
$\tilde{g}$ and areal radius $r$ have mean curvature
\begin{equation}
c = \frac{(r^2 J)_x}{3r^2}\; ,
\label{Eq:MeanExtrinsicCurv}
\end{equation}
where the function $J$ is given by $J := e^{3d} h_x/\sqrt{1-e^{4d} h_x^2}$. Here, we choose the height function such that $J=c_0 x(x^2+3b^2)/(x^2+b^2)$ with some
positive constant $c_0$. This implies that the mean curvature of the
$\tau=const$ slices is $c = c_0\left[ 1 -
\frac{2xd_x}{3}\frac{x^2+3b^2}{x^2+b^2} \right]$ which converges to
$c_0 > 0$ as $x\to\pm\infty$, and the function $e^{2d} h_x$ also satisfies the properties described above Eq. (\ref{Eq:MeanExtrinsicCurv}).

In a next step, a new, compactified space coordinate $z\in [-1,1]$ is introduced which is related to the coordinate $x\in (-\infty,+\infty)$ via the transformation $x = z/\Omega(z)$ with $\Omega\in C^\infty[-1,1]$ a smooth function which is strictly positive on $(-1,1)$, vanishes at the endpoints $z=\pm 1$, and which satisfies the inequality $L:=\Omega - z\Omega_z > 0$ everywhere on $[-1,1]$. Here, we choose $\Omega(z):=1-z^2$.

With respect to the new coordinates $(\tau,z)$ the $2$-metric assumes the form
\begin{displaymath}
\tilde{g} = \Omega^{-2}\hat{g},\qquad
\hat{g} = -\hat{\alpha}^2 d\tau^2 + \hat{\gamma}^2(dz + \hat{\beta} d\tau)^2,
\end{displaymath}
where
\begin{displaymath}
\hat{\alpha} = \sqrt{\Omega^2 e^{2d} + (\Omega J)^2},\qquad
\hat{\gamma} = \frac{L}{\hat{\alpha}}\; ,\qquad
\hat{\beta} = -\frac{\Omega J}{\hat{\gamma}}\; .
\end{displaymath}
Since $\Omega J$ is everywhere regular and positive near $z=\pm 1$,
the conformal $2$-metric $\hat{g}$ is regular for all $z\in
[-1,1]$. Since the two-dimensional wave operator is conformally
covariant, the wave equation (\ref{Eq:GeometricWaveEq}) is equivalent
to
\begin{equation}
-\hat{g}^{ab}\hat{\nabla}_a\hat{\nabla}_b u + \hat{V} u = 0,
\label{Eq:ConformalWaveEq}
\end{equation}
where $\hat{V} = \Omega^{-2}\tilde{V} = (x/z)^2\tilde{V}$ is
everywhere regular on $z\in [-1,1]$. For the numerical implementation,
we cast (\ref{Eq:ConformalWaveEq}) into first-order symmetric
hyperbolic form for the six fields $u = (u_1,u_2)$, $D = (D_1,D_2)$,
$\Pi = (\Pi_1,\Pi_2)$,
\begin{eqnarray}
u_\tau &=& \hat{\alpha}\Pi + \hat{\gamma}\hat{\beta} D,
\label{Eq:FOSH1}\\
D_\tau &=& \frac{1}{\hat{\gamma}}( \hat{\alpha}\Pi + \hat{\gamma}\hat{\beta} D )_z\; ,
\label{Eq:FOSH2}\\
\Pi_\tau &=& \frac{1}{\hat{\gamma}}( \hat{\alpha} D + \hat{\gamma}\hat{\beta}\Pi )_z
 - \hat{\alpha}\hat{V} u .
 \label{Eq:FOSH3}
\end{eqnarray}
This system is to be integrated on the compact interval $z\in
[-1,1]$. The characteristic speeds are
\begin{equation}
\lambda_\pm = \pm\frac{\hat{\alpha}}{\hat{\gamma}} + \hat{\beta}
 = \frac{1}{\hat{\gamma}}\left[ \pm \sqrt{\Omega^2 e^{2d} + (\Omega J)^2} - \Omega J \right].
\end{equation}
Therefore, $\lambda_- < 0 < \lambda_+$ on $z\in (-1,1)$. At the left
boundary, $\lambda_- = 0$ while at the right boundary, $\lambda_+ = 0$
which implies that both boundaries are outflow. This is of course
expected from the fact that the boundaries $z=\pm 1$ represent future
null infinity.

In terms of the compactified coordinates, the constraint ${\cal C} = 0$ reads
\begin{equation}
\Omega^2{\cal C} 
 = (\Omega^2 + z^2)e^{-2d}( \hat{\alpha} D_2 + \hat{\gamma}\hat{\beta}\Pi_2 )
 + \Omega^2(c_y-d_y) u_2 - \Omega^2 c_y u_1 = 0.
 \label{Eq:CompactifiedConstraint}
\end{equation}
A simple way of solving the constraints is to specify initial data for $u_2$, $\Pi_2$
and $D_2$ which are compactly supported away from the throat, and to use the
constraint ${\cal C} = 0$ and its time derivative in order to determine the remaining fields $u_1$, $\Pi_1$ and $D_1$, keeping in mind that
\begin{displaymath}
\Pi_1 = \frac{1}{\hat{\alpha}}\left( \partial_\tau u_1 - \hat{\gamma}\hat{\beta} D_1 \right),
\qquad
D_1 = \frac{1}{\hat{\gamma}}\partial_z u_1\; .
\end{displaymath}


The procedure we used to study the solution is as follows.
\begin{enumerate}
\item We set up initial data by specifying an initial pulse for $u_2$ and solving the constraint, as described above.
\item We evolve the perturbation $u$ using the system of equations 
      (\ref{Eq:FOSH1},\ref{Eq:FOSH2},\ref{Eq:FOSH3}).
\item We measure the amplitude of $u_1$ at the location of the throat.
\item We fit the resulting value of $u_1$ at the throat using the ansatz $A\cos(\omega\tau - \delta) e^{\beta\tau}$, where $\tau$ is proper time at the throat, and determine in this way the growth rate $\beta$ of the perturbation and the frequency of oscillation $\omega$ in case there is one.
\end{enumerate}

We use a finite differences approximation method with a method of lines for the evolution of the perturbation (\ref{Eq:FOSH1},\ref{Eq:FOSH2},\ref{Eq:FOSH3}). When the dimensionless charge $\nu$ is not too close to its limit value given by the constraint (\ref{Eq:WormholeParamConstr1}), second-order accurate stencils show convergence and confident results. In the case $\nu$ approaches its limit value, the results obtained from the second-order accurate stencils fail to show convergence, and we use eighth-order accurate stencils in order to reduce the errors and work in the convergence regime.

In Fig. 1 we present our results for the reflection-symmetric case $\gamma_1=0$. For small values of $\nu$ the perturbation grows exponentially and no oscillating mode shows up. However, there is a threshold of $\nu$ around $\nu^* = 0.55$ above which 
the exponential growth is modulated by a harmonic component. Also, we observe that the growth rate decreases to less than $1\%$ of the rate in the uncharged case as $\nu$ increases from zero to $0.96$. Above $\nu = 0.96$ the functions in the evolution equations become stiff and our numerical approach breaks down. In order to validate our results, we compare them with the results of the matching method described below.

An interesting question that arises from the plots in Fig. 1 is whether or not the growth rate $\beta$ may be zero for some value of $\nu$ lying between $0.965$ and $1$,  implying the existence of charged wormholes which are stable with respect to linear radial perturbations. A different possibility is that $\beta$ stays positive for all $0\leq \nu < 1$ and converges to zero for $\nu \to 1$, meaning that all supercritical wormholes are linearly unstable, but the growth rate can be made arbitrarily small by adding a sufficient amount of charge. The third possibility is that $\beta$ is bounded away from zero for all $0\leq \nu < 1$ in which case all supercritical wormholes are linearly unstable as well.
The answer to this question requires an analytic understanding of the behavior of $\beta$ as a function of $\nu$ near one and lies beyond the scope of this paper.

\begin{figure}[ht]
\includegraphics[width=8cm]{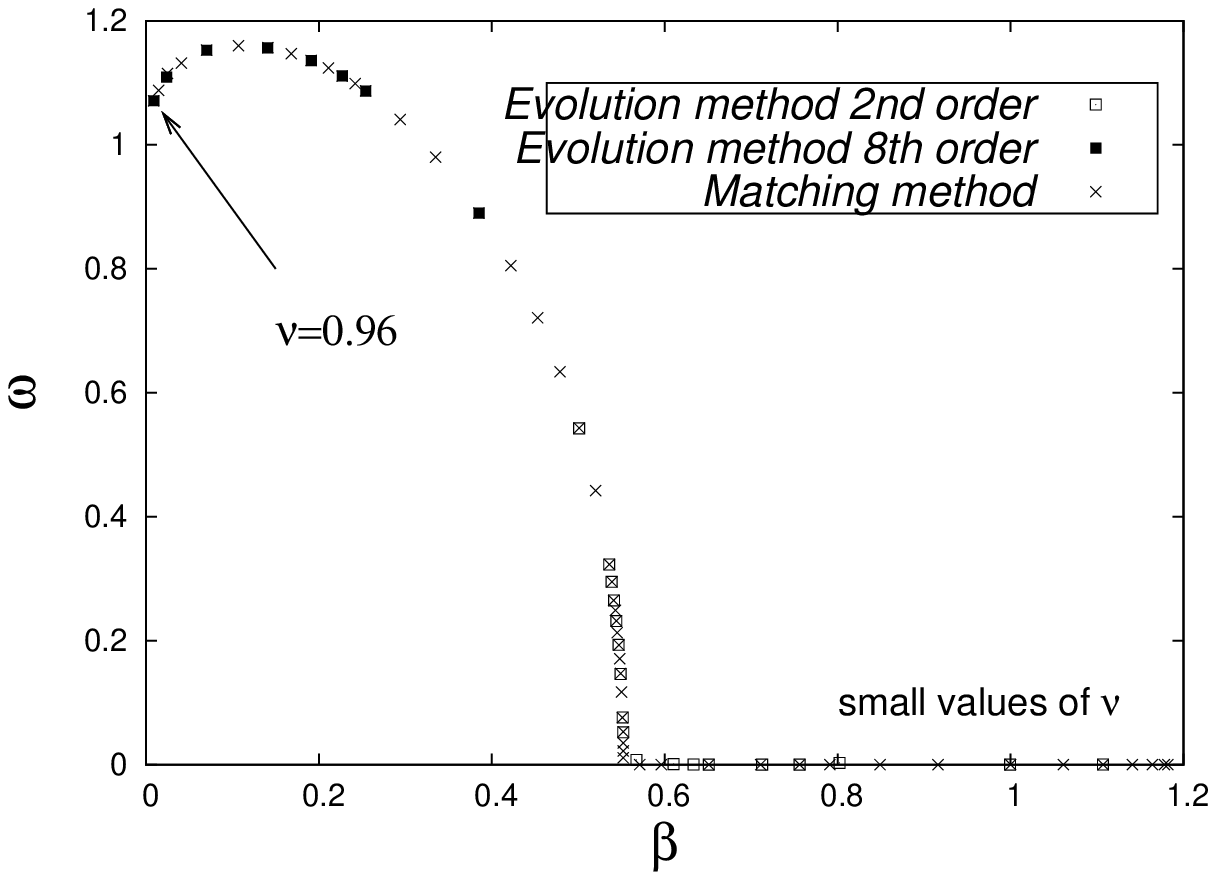}
\includegraphics[width=8cm]{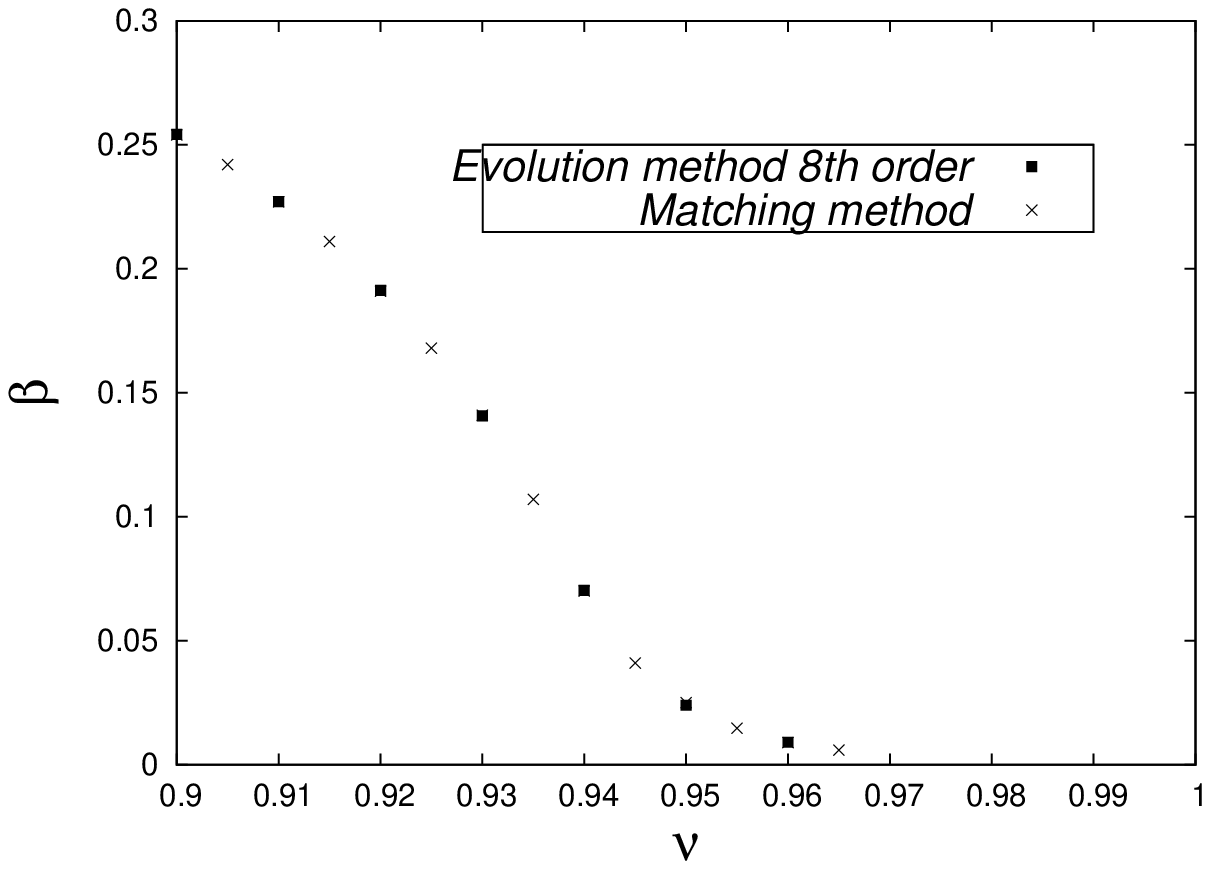}
\caption{Left panel: values of the frequency $\omega$ and growth rate $\beta$ of the perturbation for the reflection-symmetric case $\gamma_1=0$ and several values for the dimensionless charge $\nu$ using three different techniques. Each point in this plot corresponds  to a given value of $\nu$. The results in this plot indicate that the exponential growth rate decreases as the charge $\nu$ is increased. Right panel: the growth rate $\beta$ versus $\nu$ for large values of $\nu$.}
\label{fig:fig1}
\end{figure}

In Fig. 2 we present results for the asymmetric case for different values of $\gamma_1 \ne 0$. The behavior we find is similar to the massless case: for each $\gamma_1$ there is a threshold value for $\nu$ above which the perturbation shows an
oscillatory behavior while growing exponentially.

\begin{figure}[ht]
\includegraphics[width=12cm]{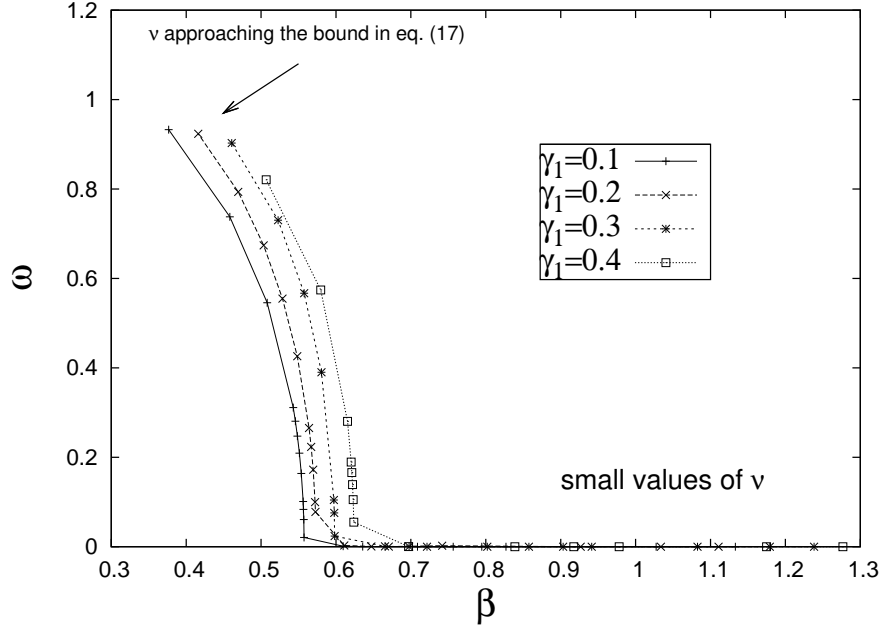}
\caption{Values of the frequency and growth rate of the perturbation for different
values for $\gamma_1$ and $\nu$. In this case the allowed values of $\nu$ are restricted by the condition (\ref{Eq:WormholeParamConstr1}) for the supercritical case. Similarly to the reflection-symmetric case there is a threshold between purely exponential growth and exponential oscillating growth.}
\label{fig:fig2}
\end{figure}


\subsection{The supercritical case: Eigenvalues of the pulsation operator}

The numerical results from the Cauchy evolution suggest the existence of eigenmodes of the pulsation operator with time-dependency $e^{(\beta + i\omega)t}$. If such an eigenmode exists, it must satisfy the constrained wave system (\ref{Eq:ConstrainedPulsation}), with the functions $u_1(t,\cdot)$ and $u_2(t,\cdot)$ being regular for all times $t \geq 0$. Defining the quantity $\Psi := e^{-c} c_y v_1 = e^c\partial_t u_2$ and using master equation I, we may rewrite the system (\ref{Eq:u1t},\ref{Eq:u2t},\ref{Eq:v1t}) in the form
\begin{eqnarray}
\partial_t u_1 &=& \frac{1}{2e^c c_{yy}}\left[ 
 \Psi_{yy} - 2d_{yy}\Psi - e^{4c}\Psi_{tt} \right],\\
\partial_t u_2 &=& e^{-c}\Psi,\\
\partial_t\Psi &=& e^{-3c}\left[ c_y(u_1)_y - (c_y d_y + 1 + \Lambda^2) u_1
 + \Lambda^2 u_2 \right],
\end{eqnarray}
where we notice the fact that $c_{yy}$ is everywhere positive, see Lemma \ref{Lem:Area} in the appendix. Since all coefficients in these equations are regular, it follows that $\Psi$ is regular if and only if $u_1$ and $u_2$ are. Therefore, we look for eigenmodes for which $\Psi$ is regular and has time-dependency $e^{(\beta + i\omega)t}$.

As a consequence of master equation I, the quantity $\Psi$ satisfies the wave equation
\begin{equation}
e^{4c}\Psi_{tt} - \Psi_{yy} + \frac{2c_{yy}}{c_y}\Psi_y 
 + 2\left( d_{yy} - \frac{d_y}{c_y} c_{yy} \right)\Psi = 0.
\end{equation}
We are looking for solutions of this equation with time-dependency $e^{(\beta + i\omega)t}$ which are everywhere regular and vanish in the asymptotic regime $y \mapsto \pm \pi/2$. This leads to the following eigenvalue problem,
\begin{equation}
 - \Psi_{yy} + \frac{2c_{yy}}{c_y}\Psi_y 
  + 2\left( d_{yy} - \frac{d_y}{c_y} c_{yy} \right)\Psi = -e^{4c}\Gamma^2\Psi,
\label{Eq:EigenvalueProblem}
\end{equation}
where $\Gamma = \beta + i\omega$. This equation has a regular singular point at the throat $c_y=0$. In the reflection-symmetric case where $\gamma_1 = 0$ one finds
\begin{eqnarray}
&& \frac{2c_{yy}}{c_y} 
 = \frac{2}{y}\left[ 1 + \frac{2}{3}(1 + \mu^2) y^2 + {\cal O}(y^4) \right],
\nonumber\\
&& 2\left( d_{yy} - \frac{d_y}{c_y} c_{yy} \right) = -\frac{4}{3}\mu^2 y^2 + {\cal O}(y^4),
\nonumber\\
&& e^{4c} = 1 + 2(1-\mu^2)y^2 + {\cal O}(y^4),
\nonumber
\end{eqnarray}
and we obtain a reflection-symmetric, local solution of the form
\begin{equation}
\Psi_1(y) = 1 - \frac{1}{2}\Gamma^2 y^2
 - \frac{1}{24}\left[ 3\Gamma^4 + 4\Gamma^2(5\mu^2 - 1) + 8\mu^2 \right] y^4
 + {\cal O}(y^6).
\label{Eq:LocalSolutionThroat}
\end{equation}
In order to understand the asymptotic behavior of the solutions of Eq. (\ref{Eq:EigenvalueProblem}) when $y\to \pm \pi/2$ approaches the asymptotic regime, it is easier to go back to the consideration of master equation I, which in terms of the coordinate $\rho$ defined in (\ref{Eq:rhoDef}) reads
\begin{equation}
-\partial_\rho^2 v_1 + \left[ \Gamma^2 + \hat{V}_{11}(\rho) \right] v_1 = 0,
\end{equation}
where we have assumed again a time-dependency $e^{\Gamma t}$ for $v_1$.
In the reflection-symmetric case, we find
\begin{displaymath}
\hat{V}_{11} = \frac{2\mu\tan\left(\mu\frac{\pi}{2}\right)}{\cos^4\left(\mu\frac{\pi}{2}\right)}
\frac{1}{x^3}\left[ 1 + {\cal O}\left( \frac{1}{x} \right) \right].
\end{displaymath}
According to standard theorems\footnote{See, for instance, Ref. \cite{Hille-Book}.}, there exist local solutions near $\rho=+\infty$ such that
\begin{displaymath}
v_1 = e^{-\Gamma\rho}\left[ 1 + h(\rho) \right],
\end{displaymath}
where $h$ is a $C^2$-function satisfying $h(\rho)\to 0$, $h_\rho(\rho)\to 0$ when $\rho\to+\infty$. Therefore, we obtain the asymptotic solution
\begin{equation}
\Psi_2 = e^{-c} c_y e^{-\Gamma\rho}\left[ 1 + h(\rho) \right]
\label{Eq:LocalSolutionAsymptotic}
\end{equation}
for $y \to \pi/2$. Our numerical method consists in integrating the local solutions (\ref{Eq:LocalSolutionThroat},\ref{Eq:LocalSolutionAsymptotic}) and to match their Wronski determinant
\begin{displaymath}
W(y) = \det\left( \begin{array}{rr} 
 \Psi_1(y) & \Psi_2(y) \\
 \partial_y\Psi_1(y) & \partial_y\Psi_2(y)
 \end{array} \right)
\end{displaymath}
at some intermediate point $0 < y_1 < \pi/2$ by fine-tuning the complex eigenvalue $\Gamma$. The result is shown in Fig. 1, and agrees very well (within $5\%$ of relative error) with the growth rate and oscillation frequency found from the Cauchy evolution.

\section{Conclusions}
\label{Sec:Conclusions}

In this article we analyze the stability of static, spherically symmetric general relativistic wormhole solutions sourced by a massless ghost scalar field and an electromagnetic field. To this purpose we first construct the complete spectrum of such solutions. Among the solutions we distinguish between three types depending on the values of the charge and mass parameters: subcritical, critical and supercritical. We show that in the first two cases all solutions are unstable with respect to linear spherically symmetric perturbations. This is done by reducing the perturbation equation to a Sturm-Liouville problem and using standard tools in Schr\"odinger operator theory.

In the supercritical case we are not able to reduce the perturbation equations to a
scalar equation which is everywhere regular. We instead obtain a constrained wave system for two gauge-invariant quantities. We study this system numerically as a Cauchy evolution problem based on a domain compactification scheme and observe the growth of the perturbations at the wormhole throat. The analysis reveals: 
i) for small values of the dimensionless charge parameter $\nu$ the growth of the perturbation is exponential in proper time, 
ii) there is a threshold $\nu^{*}$ in the value of the charge above which the exponential growth is modulated by an oscillatory component, 
iii) the growth rate decreases when the value of the charge increases, 
iv) as the value of the charge approaches its limiting value the functions in the evolution equations become stiff and our numerical approach breaks down. However, we find consistent results up to values of $\nu = 0.96$. Within this regime we can decrease the growth rate by a factor of more than $100$ compared to the uncharged case. To validate our numerical results we also analyzed the eigenvalues of the spatial operator in the perturbation equation. Based on a matching method we find a real eigenvalue for $\nu < \nu^{*}$. Above that value the eigenvalue acquires a nontrivial imaginary part, explaining the oscillatory behavior found in the Cauchy evolution. The eigenvalues fit the growth rate and the oscillation frequency obtained with the Cauchy evolution within an accuracy of $5\%$. The fact that we find complex eigenvalues provides an explanation for the inability to reduce the perturbation equations to a Sturm-Liouville problem, for which all eigenvalues are necessarily real.

The motivation for this work was to study a possible mechanism for stabilizing the static, spherically symmetric wormholes solutions supported by a ghost scalar field. We have analyzed the stability behavior when such solutions are charged up by an electromagnetic field. The fact that this mechanism does not seem to be able to stabilize the wormholes raises some doubts about the success of other similar mechanisms, like adding angular momentum.

\acknowledgments

It is a pleasure to thank H. Beyer, A. Merzon, U. Nucamendi and T. Zannias for stimulating discussions. This work was supported in part by grants 
CIC 4.9, 4.19 and 4.23 to Universidad Michoacana, 
PROMEP UMICH-PTC-195, UMICH-PTC-210 and UMICH-CA-22 from SEP Mexico and 
CONACyT grants 61173, 79601 and 79995.

\appendix
\section{Some technical properties of the static wormhole solutions}
\label{Sec:App}

The family of charged, static wormhole metrics is determined by the function
\begin{displaymath}
d = \left\{ \begin{array}{ll}
\gamma_0 - \log\left[ \cosh(\Lambda y) - \gamma_1\frac{\sinh(\Lambda y)}{\Lambda} \right] &, \Lambda > 0,\\
\gamma_0 - \log\left[ 1 - \gamma_1 y \right] &, \Lambda = 0,\\
\gamma_0 - \log\left[ \cos(\mu y) - \gamma_1\frac{\sin(\mu y)}{\mu} \right] &, \Lambda = i\mu, 0 < \mu < 1,
\end{array}
\right.
\end{displaymath}
where $\Lambda := \sqrt{\gamma_1^2 - e^{2\gamma_0}\epsilon^2}$ and $y=\arctan(x/b)$. Here, $\gamma_1$ is subject to the inequalities (\ref{Eq:WormholeParamConstr1}).
For simplicity, we choose $b=1$ and $\gamma_0=0$ in what follows. Then, the metric functions $d$ and $c = \log\sqrt{x^2+1} - d$ satisfy the following relations,
\begin{equation}
d_{yy} = \epsilon^2 e^{2d}, \qquad
c_{yy} = 1 + x^2 - \epsilon^2 e^{2d}, \qquad
(2d_y+c_y)c_y = x^2 - \epsilon^2 e^{2d} - \Lambda^2,
\label{Eq:BackgroundA1}
\end{equation}
which follow from Eqs. (\ref{Eq:Static1},\ref{Eq:Static2},\ref{Eq:Static4}) with $e^{2(c+d)} = 1 + x^2$, $\epsilon^2 = \kappa_0 Q^2/2$ and the relation (\ref{Eq:WormholeParamConstr2}). This together with $c_y = x - d_y$ also implies the equations
\begin{equation}
d_y^2 = \Lambda^2 + \epsilon^2 e^{2d}, \qquad
d_{yy} = d_y^2 - \Lambda^2, \qquad
c_{yy} = 1 + \Lambda^2 + x^2 - d_y^2
\label{Eq:BackgroundA2}
\end{equation}
which turn out to be useful. The next result is related to the global behavior of the areal radius $r = e^{c}$.

\begin{lemma}
\label{Lem:Area}
The function $c$ has a unique local minimum and $c_{yy}$ is strictly positive on the interval $-\infty < x < +\infty$.
\end{lemma}

\proof
First, we notice from Eq. (\ref{Eq:BackgroundA2}) that at points where $c_y=0$ we must have $c_{yy} = 1 + \Lambda^2 > 0$ which shows that local extrema of $c$ are necessarily local minima. Since $c\to+\infty$ when $x\mapsto \pm\infty$ it follows that $c$ has a unique local minimum at some point $x = x_{throat}$.

Next, we prove that $c_{yy}$ is everywhere positive. For this, we first notice that in the uncharged case $\epsilon=0$, it follows from Eq. (\ref{Eq:BackgroundA1}) that $c_{yy} = 1 + x^2 \geq 1$. Next, assume that $\varepsilon > 0$ and that $c_{yy}$ has a zero at some point $x^*$. Since $c_{yy} = 1 + \Lambda^2 > 0$ at $x_{throat}$, $x^*$ lies either to the left or to the right of $x_{throat}$. Suppose that $x^* < x_{throat}$. Then, from Eq. (\ref{Eq:BackgroundA2}), it follows that $c_{yyy} = 2d_{yy} c_y < 0$ at $x=x^*$, where we have used $d_{yy} = \epsilon^2 e^{2d} > 0$ and $c_y < 0$ for points to the left of the throat. This means that $c_{yy}$ crosses the $x$-axis from above at all its zeroes lying to the left of $x_{throat}$. However, since $c_{yy} > 0$ at $x_{throat}$, this means that $c_{yy}$ cannot have zeroes to the left of $x_{throat}$. A similar argument shows that $c_{yy}$ cannot have zeroes to the right of $x_{throat}$ either.
\qed

Next, we analyze the properties of the function
\begin{displaymath}
F = 1 + xy - \frac{y c_y}{1+\Lambda^2}
  = 1 + \frac{(\Lambda^2 x + d_y)y}{1+\Lambda^2}
\end{displaymath}
defined in (\ref{Eq:DefF}). Using Eqs. (\ref{Eq:BackgroundA2}) it is not difficult to show that $F$ satisfies the relation
\begin{equation}
F_y = d_y F + (1+xy)\frac{\Lambda^2}{1+\Lambda^2} c_y\; .
\label{Eq:FDeriv}
\end{equation}
For the linear stability analysis, it is important to know whether or not $F$ has zeroes.

\begin{lemma}
\label{Lem:F}
In the subcritical and critical cases, the function $F$ is strictly positive on the interval $-\infty < x < +\infty$. In the supercritical case, $F$ has precisely two zeroes.
\end{lemma}

\proof
In the critical case, $F = (1 - \gamma_1 y)^{-1}$ which is everywhere positive. In the other cases, since $d_y$ is bounded and $\Lambda^2\neq 0$, we see from its definition that $F$ satisfies $F\to+\infty$ ($F\to -\infty$) as $|x|\to+\infty$ in the subcritical (supercritical) case. On the other hand, at $x_{throat}$ where $c_y=0$, we have $F = 1 + xy > 0$. Finally, let $x^*$ be a zero of $F$ which lies to the left of $x_{throat}$. From Eq. (\ref{Eq:FDeriv}) it follows that $(1 + \Lambda^2) F_y = (1 + xy)\Lambda^2 c_y$ at $x^*$ which is negative (positive) in the subcritical (supercritical) case. This means that all zeroes of $F$ lying to the left of $x_{throat}$ cross the $x$-axis from above (below). A similar argument shows that all zeroes of $F$ lying to the right of $x_{throat}$ cross the $x$-axis from below (above). As a consequence, $F$ has no zeroes in the subcritical case and two zeroes (one to the left and one to the right of $x_{throat}$) in the supercritical case.
\qed

\bibliographystyle{unsrt}
\bibliography{refs}

\begin{thebibliography}{10}

\bibitem{mMkTyU88}
M.S. Morris, K.S. Thorne, and U.~Yurtsever.
\newblock Wormholes, time machines, and the weak energy condition.
\newblock {\em Phys. Rev. Lett.}, 61:1446--1449, 1988.

\bibitem{mMkT88}
M.S. Morris and K.S. Thorne.
\newblock Wormholes in spacetime and their use for interstellar travel: {A}
  tool for teaching general relativity.
\newblock {\em Am. J. Phys.}, 56:395--412, 1988.

\bibitem{vFiN90}
V.P. Frolov and I.D. Novikov.
\newblock Physical effects in wormholes and time machines.
\newblock {\em Phys. Rev. D}, 42:1057--1065, 1990.

\bibitem{tDsS07}
T.~Damour and S.N. Solodukhin.
\newblock Wormholes as black hole foils.
\newblock {\em Phys. Rev. D}, 76:024016, 2007.

\bibitem{nMtZ08}
N.~Montelongo Garc\'{\i}a and T.~Zannias.
\newblock Structure of the effective potential for a spherical wormhole.
\newblock {\em Phys. Rev. D}, 78:064003, 2008.

\bibitem{jFkSdW93}
J.L. Friedman, K.~Schleich, and D.M. Witt.
\newblock Topological censorship.
\newblock {\em Phys. Rev. Lett.}, 71:1486--1489, 1993.

\bibitem{tRjW57}
T.~Regge and J.~Wheeler.
\newblock Stability of a {S}chwarzschild singularity.
\newblock {\em Phys. Rev.}, 108:1063--1069, 1957.

\bibitem{fZ70}
F.~Zerilli.
\newblock Effective potential for even-parity {R}egge-{W}heeler gravitational
  perturbation equations.
\newblock {\em Phys. Rev. Lett.}, 24:737--738, 1970.

\bibitem{bKrW87}
B.~S. Kay and R.~M. Wald.
\newblock Linear stability of {S}chwarzschild under perturbations which are
  non-vanishing on the bifurcation $2$-sphere.
\newblock {\em Class. Quantum Grav.}, 4:893--898, 1987.

\bibitem{vM74a}
V.~Moncrief.
\newblock Odd-parity stability of a {R}eissner-{Nordstr\"om} black hole.
\newblock {\em Phys. Rev. D}, 9:2707--2709, 1974.

\bibitem{vM74b}
V.~Moncrief.
\newblock Stability of a {R}eissner-{Nordstr\"om} black holes.
\newblock {\em Phys. Rev. D}, 10:1057--1059, 1974.

\bibitem{vM75}
V.~Moncrief.
\newblock Gauge-invariant perturbations of {R}eissner-{Nordstr\"om} black
  holes.
\newblock {\em Phys. Rev. D}, 12:1526--1537, 1975.

\bibitem{jGfGoS09a}
J.~A. Gonz\'alez, F.~S. Guzm\'an, and O.~Sarbach.
\newblock Instability of wormholes supported by a ghost scalar field. {I}.
  {L}inear stability analysis.
\newblock {\em Class. Quantum Grav.}, 26:015010, 2009.

\bibitem{jGfGoS09b}
J.~A. Gonz\'alez, F.~S. Guzm\'an, and O.~Sarbach.
\newblock Instability of wormholes supported by a ghost scalar field. {II}.
  {N}onlinear evolution.
\newblock {\em Class. Quantum Grav.}, 26:015011, 2009.

\bibitem{pKsS08}
P.E. Kashargin and S.V. Sushkov.
\newblock Slowly rotating wormholes: The first order approximation.
\newblock {\em Grav. Cosmol.}, 14:80--85, 2008.

\bibitem{tMdN06}
T.~Matos and D.~N\'u{\~n}ez.
\newblock Rotating scalar field wormhole.
\newblock {\em Class. Quant. Grav.}, 23:4485--4496, 2006.

\bibitem{tM09}
T.~Matos.
\newblock Class of {E}instein-{M}axwell phantom fields: Rotating and magnetised
  wormholes.
\newblock arxiv/0902.4439 [gr-qc].

\bibitem{hE73}
H.G. Ellis.
\newblock Ether flow through a drainhole: {A} particle model in general
  relativity.
\newblock {\em J. Math. Phys.}, 14:104--118, 1973.

\bibitem{kB73}
K.A. Bronnikov.
\newblock Scalar-tensor theory and scalar charge.
\newblock {\em Acta Phys. Polonica B}, 4:251--266, 1973.

\bibitem{cMdS64}
C.W. Misner and D.H. Sharp.
\newblock Relativistic equations for adiabatic, spherically symmetric
  gravitational collapse.
\newblock {\em Phys. Rev.}, 136:B571--B576, 1964.

\bibitem{oBmHnS96}
O.~Brodbeck, M.~Heusler, and N.~Straumann.
\newblock Pulsations of spherically symmetric systems in general relativity.
\newblock {\em Phys. Rev. D}, 53:754--761, 1996.

\bibitem{CourantHilbert-Book}
R.~Courant and D.~Hilbert.
\newblock {\em Methods of Mathematical Physics, Volume I}.
\newblock Wiley Classics Edition, New-York, 1989.

\bibitem{hApQ95}
H.~Amann and P.~Quittner.
\newblock A nodal theorem for coupled systems of {Schr\"odinger} equations and
  the number of bound states.
\newblock {\em J. Math. Phys.}, 36:4553--4560, 1995.

\bibitem{aZdNsH09}
A.~Zengino\u{g}lu, D.~N\'u{\~n}ez, and S.~Husa.
\newblock Gravitational perturbations of {S}chwarzschild spacetime at null
  infinity and the hyperboloidal initial value problem.
\newblock {\em Class. Quantum Grav.}, 26:035009, 2009.

\bibitem{Hille-Book}
E.~Hille.
\newblock {\em Lectures on Ordinary Differential Equations}.
\newblock Addison-Wesley, Reading, 1969.

\end{thebibliography}

\end{document}